\documentclass[preprintnumbers,twocolumn,showpacs,amsmath,amssymb]{revtex4}
\usepackage{epsfig}% Include figure files
\usepackage{graphicx}% Include figure files
\usepackage{dcolumn}% Align table columns on decimal point
\usepackage{bm}% bold math

\newcommand{\be}{\begin{equation}}
\newcommand{\ee}{\end{equation}}

\newcommand{\bea}{\begin{eqnarray}}
\newcommand{\eea}{\end{eqnarray}}
\newcommand{\bfig}{\begin{figure}}
\newcommand{\efig}{\end{figure}}
\newcommand{\bc}{\begin{center}}
\newcommand{\ec}{\end{center}}

\begin{document}
\preprint{ZU-TH 16/07, IPPP-07-36, Edinburgh 2007/10, HEPTOOLS 07-016}

\title{Second-order QCD corrections to the thrust distribution}

\author{A.\ Gehrmann-De Ridder$^a$, T.\ Gehrmann$^b$, E.W.N.\ Glover$^c$,
G.\ Heinrich$^d$}
 \affiliation{$^a$ Institute for Theoretical Physics, ETH, CH-8093 Z\"urich,
Switzerland\\
$^b$ Institut f\"ur Theoretische Physik,
Universit\"at Z\"urich, CH-8057 Z\"urich, Switzerland\\
$^c$ Institute of Particle Physics Phenomenology, 
        Department of Physics,
        University of Durham, Durham, DH1 3LE, UK\\
$^d$ School of Physics, The University of Edinburgh, Edinburgh EH9 3JZ,
Scotland}

\date{\today}% It is always \today, today,

\begin{abstract}
We compute the next-to-next-to-leading order (NNLO) 
QCD corrections to the 
thrust distribution in electron-positron annihilation. The corrections 
turn out to be sizable, enhancing the previously known next-to-leading 
order prediction by about 15\%. Inclusion of the NNLO corrections 
significantly reduces
the theoretical renormalisation scale uncertainty on the prediction of the 
thrust distribution.
\end{abstract}

%Use showkeys class option if keyword
\maketitle

%%%%%%%%%%%%%%%%%%%%%% 72 characters %%%%%%%%%%%%%%%%%%%%%%%%%%%%%%%%%%%
Three-jet production cross sections and related event shape distributions
in $e^+e^-$ annihilation processes 
are classical hadronic observables which can be measured very accurately
and provide an ideal proving ground for testing our understanding of strong
interactions.
The deviation from simple two-jet configurations is proportional
to the strong coupling constant, so that 
by comparing the measured three-jet rate and related 
event shapes  with the theoretical 
predictions, one can determine the strong coupling constant 
$\alpha_{s}$.  

The theoretical prediction is made within perturbative QCD, expanded to a
finite order in the coupling constant. This truncation of the perturbative 
series induces a theoretical uncertainty from omitting higher order terms.
It can be quantified by the renormalisation scale dependence of the 
prediction, which is vanishing for an all-order prediction. The residual 
dependence on variations of the renormalisation scale is therefore an estimate 
of the theoretical error.

So far the three-jet rate and related event shapes
have been calculated~\cite{ert,ert2} up to the next-to-leading order (NLO),
improved by a resummation of leading and subleading infrared 
logarithms~\cite{ctwt,ctw} and by the inclusion of power
corrections~\cite{power}. 

QCD studies of event shape 
observables at LEP~\cite{expreview} are based around the use of 
NLO parton-level event generator 
programs~\cite{event}.  
It turns out that the current error on $\alpha_s$ from these observables  
\cite{Bethke} is dominated by the theoretical uncertainty.
Clearly, to improve the determination of $\alpha_{s}$, 
the calculation of the NNLO corrections to these
observables becomes mandatory. We present here the first NNLO calculation 
of the thrust distribution, 
which is an event shape related to three-jet production. 

The thrust variable for a hadronic final state in $e^+e^-$ annihilation is 
defined as~\cite{farhi} 
\begin{displaymath}
T=\max_{\vec{n}}
\left(\frac{\sum_i |\vec{p_i}\cdot \vec{n}|}{\sum_i |\vec{p_i}|}\right)\,,
\end{displaymath}
where $p_i$ denotes the three-momentum of particle $i$, with the sum running 
over all particles. The unit vector $\vec{n}$ is varied to find  the 
thrust direction $\vec{n}_T$ which maximises the expression in parentheses 
on the right hand side. 

It can be seen that a two-particle final state has fixed $T=1$,
consequently the thrust distribution receives its first non-trivial 
contribution from three-particle final states, which, at order $\alpha_s$, 
correspond to three-parton final states. Therefore, 
both theoretically and experimentally, the thrust distribution is 
closely related to three-jet production. 

Three-jet production at tree-level is induced by the decay of a virtual
photon (or other neutral gauge boson) into a quark-antiquark-gluon final
state. At higher orders, this process receives corrections from extra
real or virtual particles. The individual partonic channels
that contribute through to NNLO
are shown in Table~\ref{table:partons}. All of the tree-level and loop
amplitudes associated with these channels are known 
in the literature~\cite{3jme,muw2,V4p,tree5p}.

\begin{table}[t]
\begin{tabular}{lll}
\hline\\
LO & $\gamma^*\to q\,\bar qg$ & tree level \\[2mm]
NLO & $\gamma^*\to q\,\bar qg$ & one loop \\
 & $\gamma^*\to q\,\bar q\, gg$ & tree level \\
 & $\gamma^*\to q\,\bar q\, q\bar q$ & tree level \\[2mm]
NNLO & $\gamma^*\to q\,\bar qg$ & two loop \\
 & $\gamma^*\to q\,\bar q\, gg$ & one loop \\
& $\gamma^*\to q\,\bar q\, q\,\bar q$ & one loop \\
& $\gamma^*\to q\,\bar q\, q\,\bar q\, g$ & tree level \\
& $\gamma^*\to q\,\bar q\, g\,g\,g$ & tree level\\[2mm]
\hline
\end{tabular}
\caption{Partonic contributions to the thrust distribution in 
perturbative QCD.\label{table:partons}}
\end{table}

For a given partonic final state, thrust is computed according to 
the same definition as in the experiment, which is applied to partons instead 
of hadrons. At leading order, all three final state partons must be 
well separated from each other, to allow $T$ to differ from the trivial 
two-parton limit. At NLO, up to four partons 
can be present in the final state, 
two of which 
can be clustered together,   
whereas at NNLO, the final state can consist of up to five partons, 
such that as many as three partons can be clustered together. 
The more partons in the final state, 
the better one expects the matching between theory and 
experiment to be.

The two-loop $\gamma^* \to q\bar q g$ matrix elements were derived 
in~\cite{3jme} by reducing all relevant Feynman integrals to a small 
set of master integrals using integration-by-parts~\cite{ibp} and 
Lorentz invariance~\cite{gr} identities, solved with the Laporta 
algorithm~\cite{laporta}. The master integrals~\cite{3jmaster} were 
computed from their differential equations~\cite{gr} and expressed 
analytically
in terms of one- and two-dimensional harmonic polylogarithms~\cite{hpl}. 

The one-loop four-parton matrix elements relevant here~\cite{V4p} were 
originally derived in the context of NLO corrections to four-jet 
production and related event shapes~\cite{fourjetprog,cullen}. One of 
these four-jet parton-level event 
generator programs~\cite{cullen} is the starting point 
for our calculation, since it already contains all relevant 
four-parton and five-parton matrix elements.

The four-parton and five-parton contributions to three-jet-like final 
states at NNLO contain infrared real radiation singularities, which 
have to be extracted and combined with the 
infrared singularities~\cite{catani} 
present in the virtual three-parton and four-parton contributions to 
yield a finite result. In our case, this is accomplished by 
introducing subtraction functions, which account for the 
infrared real radiation singularities, and are sufficiently simple to 
be integrated analytically. Schematically, this subtraction reads:
\begin{eqnarray*}
\lefteqn{{\rm d}\sigma_{NNLO}=\int_{{\rm d}\Phi_{5}}\left({\rm d}\sigma^{R}_{NNLO}
-{\rm d}\sigma^{S}_{NNLO}\right) }\nonumber \\ 
&&+\int_{{\rm d}\Phi_{4}}\left({\rm d}\sigma^{V,1}_{NNLO}
-{\rm d}\sigma^{VS,1}_{NNLO}\right)\nonumber \\&&
+ \int_{{\rm d}\Phi_{5}}
{\rm d}\sigma^{S}_{NNLO}
+\int_{{\rm d}\Phi_{4}}{\rm d}\sigma^{VS,1}_{NNLO} 
% \nonumber \\ &&
+ \int_{{\rm d}\Phi_{3}}{\rm d}\sigma^{V,2}_{NNLO}\;,
\end{eqnarray*}
where ${\rm d} \sigma^{S}_{NNLO}$ denotes the real radiation subtraction term 
coinciding with the five-parton tree level cross section 
 ${\rm d} \sigma^{R}_{NNLO}$ in all singular limits~\cite{doubleun}. 
Likewise, ${\rm d} \sigma^{VS,1}_{NNLO}$
is the one-loop virtual subtraction term 
coinciding with the one-loop four-parton cross section 
 ${\rm d} \sigma^{V,1}_{NNLO}$ in all singular limits~\cite{onelstr}. 
Finally, the two-loop correction 
to the three-parton cross section is denoted by ${\rm d}\sigma^{V,2}_{NNLO}$.
With these, each line in the above equation is individually 
infrared finite, and 
can be integrated numerically.

Systematic methods to derive and integrate subtraction terms 
were available in the literature only to NLO~\cite{nlosub,ant},
with extension to NNLO in special cases~\cite{cshiggs}.
In the context of 
this project, we fully developed an NNLO subtraction formalism~\cite{ourant}, 
based on the antenna subtraction method originally proposed at 
NLO~\cite{cullen,ant}. 
The basic idea of the antenna subtraction approach is to construct 
the subtraction terms  from antenna functions. 
Each antenna function encapsulates 
all singular limits due to the 
 emission of one or two unresolved partons between two colour-connected hard
partons.
This construction exploits the universal factorisation of 
phase space and squared matrix elements in all unresolved limits.
The individual antenna functions are obtained by normalising 
three-parton and four-parton tree-level matrix elements and 
three-parton one-loop matrix elements 
to the corresponding two-parton 
tree-level matrix elements. Three different types of 
antenna functions are required,
corresponding to the different pairs of hard partons 
forming the antenna: quark-antiquark, quark-gluon and gluon-gluon antenna 
functions. All these can be derived systematically from matrix 
elements~\cite{our2j} for physical processes. 

The factorisation of the final state phase space into antenna phase 
space and hard phase space requires a mapping of the antenna momenta 
onto reduced hard momenta. We use the mapping derived in~\cite{dak1} for 
the three-parton and four-parton antenna functions. To extract the infrared 
poles of the subtraction terms, the antenna functions must be integrated 
analytically over the appropriate antenna phase spaces, which is done by 
reduction~\cite{babis} to known 
phase space master integrals~\cite{ggh}. 

We tested the proper implementation of 
the subtraction by generating trajectories of phase space points approaching 
a given single or double unresolved limit. 
Along these trajectories, we observe that the 
antenna subtraction terms converge towards the physical matrix 
elements, and that the cancellations among individual 
contributions to the subtraction terms take place as expected. 
Moreover, we checked the correctness of the 
subtraction by introducing a 
lower cut (slicing parameter) on the phase space variables, and observing 
that our results are independent of this cut (provided it is 
chosen small enough). This behaviour indicates that the 
subtraction terms ensure that the contribution of potentially singular 
regions of the final state phase space does not contribute to the numerical 
integrals, but is accounted for analytically. A detailed description of the 
calculation will be given elsewhere~\cite{our3j}. 

The resulting numerical programme, {\tt EERAD3}, yields the full kinematical 
information on a given multi-parton final state. It 
can thus be used to compute any 
infrared-safe observable related to three-particle final states at 
${\cal O}(\alpha_s^3)$. As a first application, we present results for the 
NNLO corrections to the thrust distribution here. Leaving aside numerically 
small pure singlet-corrections (which come from $\gamma^*\to ggg$ and 
related final states, and appear first at NNLO), the theoretical 
expression for the thrust distribution through to NNLO can be expressed 
by three dimensionless coefficients ($A,B,C$), 
which depend only on $T$ and not on the scale 
of the process or on coupling constants and quark charges. These are obtained 
as coefficients of the thrust distribution normalised to the tree-level 
cross section $\sigma_0$ for $e^+e^- \to q\bar q$, evaluated for 
$\alpha_s=\alpha_s(Q)$, where $Q$ is the centre-of-mass energy of the process:
\begin{displaymath}
\frac{1}{\sigma_{0}}\, \frac{{\rm d}\sigma}{{\rm d} T} = 
\left(\frac{\alpha_s}{2\pi}\right) \frac{{\rm d} A}{{\rm d} T} 
 +  \left(\frac{\alpha_s}{2\pi}\right)^2 \frac{{\rm d} B}{{\rm d} T} 
+ \left(\frac{\alpha_s}{2\pi}\right)^3 \frac{{\rm d} C}{{\rm d} T}\,.
\end{displaymath}
The experimentally measured thrust distribution 
\begin{displaymath}
\frac{1}{\sigma_{{\rm had}}}\, \frac{{\rm d}\sigma}{{\rm d} T}
\end{displaymath}
can then be expressed in terms of $A,B,C$ by expanding $\sigma_{{\rm had}}$ 
around $\sigma_0$. If the renormalisation scale of $\alpha_s$ is chosen 
to be $\mu \neq Q$, additional terms proportional to powers of 
$\ln (\mu^2/Q^2)$ appear, which are again expressed in terms of $A,B,C$
and of the coefficients of the QCD $\beta$-function. 
\begin{figure}[t]
\begin{center}
\epsfig{file=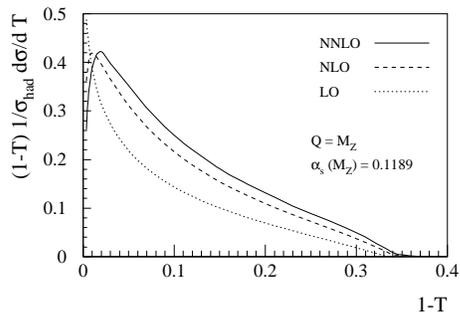,angle=-90,width=6cm}
\end{center}
\caption{Thrust distribution at $Q= M_Z$.\label{fig:thrust}}
\end{figure}

In the numerical evaluation, we use $M_Z= 91.1876$~GeV and $\alpha_s(M_Z)=
0.1189$~\cite{Bethke}. Figure~\ref{fig:thrust} displays the perturbative 
expression for the thrust distribution at LO, NLO and NNLO, evaluated 
for $\mu =Q = M_Z$. It can be seen that inclusion of the NNLO corrections 
enhances the thrust distribution by around (15-20)\% over the 
 range $0.03 < (1-T) < 0.33$. Outside this range, one does not expect the 
perturbative fixed-order prediction to yield reliable results. For 
$(1-T)>0.33$,
the leading-order prediction vanishes due to kinematical constraints from 
having only three partons in the final state; at NLO, $(1-T)>0.42$ is 
kinematically forbidden, 
and the spherical 
limit $T\to 0.5$ is reached only for infinitely many partons in 
the final state. For $T\to 0$, the convergence of the perturbative series 
is spoilt by powers of
logarithms $\ln(1-T)$ appearing in higher perturbative orders, 
thus necessitating an all-order resummation of these logarithmic 
terms~\cite{ctwt,ctw}, and a matching of fixed-order and resummed 
predictions~\cite{hasko}.
\begin{figure}[t]
\begin{center}
\epsfig{file=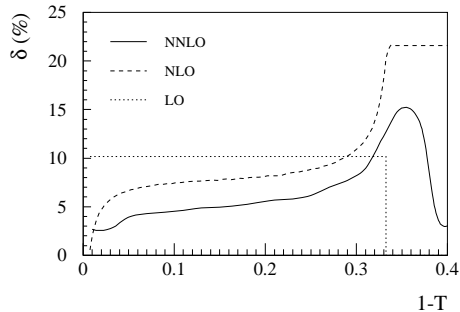,angle=-90,width=6cm}
\end{center}
\caption{Relative scale-uncertainty on thrust distribution at 
different orders in perturbation theory.\label{fig:thrusterr}}
\end{figure}

To estimate the theoretical error inherent to the perturbative prediction,
we vary the renormalisation scale in the interval $\mu \in [M_Z/2;2\, M_Z]$.
The relative uncertainty at each order 
\begin{displaymath}
\delta = \frac{\max_{\mu} (\sigma(\mu)) - \min_\mu (\sigma(\mu)) }
{2 \sigma (\mu = M_Z)}
\end{displaymath}
is displayed in Figure~\ref{fig:thrusterr}. It can be clearly seen how  
the inclusion of higher-order corrections stabilises the prediction, 
and the uncertainty $\delta$ is reduced by about 30\% 
between NLO and NNLO. The increase in uncertainty above $1-T=0.33$ is
 due to  the vanishing of the leading order contribution; the perturbative 
fixed-order description and its theoretical error become unreliable beyond 
this point.

\begin{figure}[b]
\begin{center}
\epsfig{file=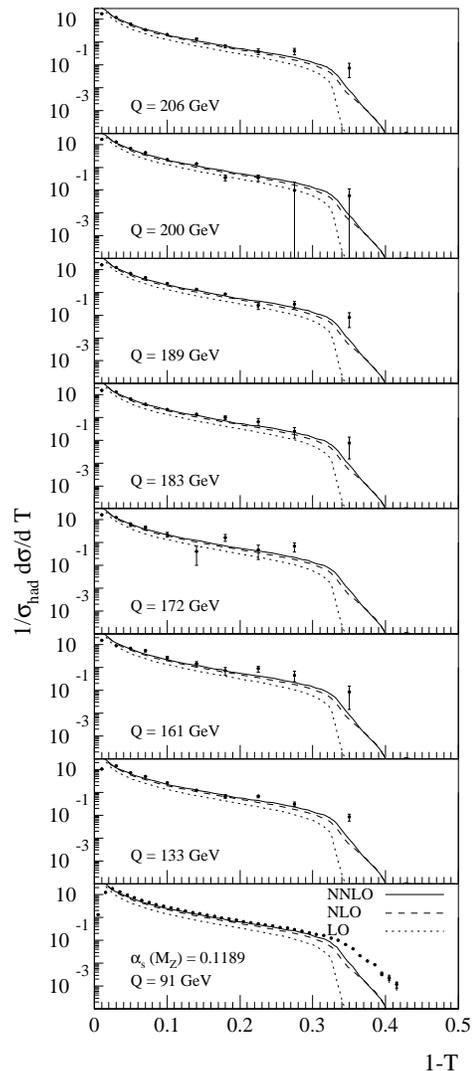,angle=-90,width=6cm}
\end{center}
\caption{NNLO thrust distribution compared to experimental 
data from ALEPH~\protect{\cite{aleph}}.\label{fig:thrustexp}}
\end{figure}
In Figure~\ref{fig:thrustexp}, we compare  the theoretical NNLO prediction
for the thrust distribution to experimental data from the ALEPH 
experiment~\cite{aleph}. 
Similar measurements were carried out by all LEP experiments~\cite{lep}. 
The NNLO correction is positive and
for the same value of $\alpha_s$, yields a prediction that is larger than
at NLO. This indicates the need 
for an improved  determination of $\alpha_s$ from event shape
 data, which takes the 
newly computed NNLO corrections into account. Work on this is ongoing. 

In this letter, we presented the first NNLO calculation of event shapes 
related to three-jet production in $e^+e^-$ annihilation. We developed 
a numerical programme which can compute any infrared-safe observable
through to ${\cal O}(\alpha_s^3)$, which we applied here to determine the 
NNLO corrections to the thrust distribution. These corrections are moderate,
indicating the convergence of the perturbative expansion. Their inclusion 
results in a considerable reduction of the theoretical error on the 
thrust distribution. Our results will allow a 
significantly  improved determination of 
the strong coupling constant from jet observables.

{\bf Acknowledgement:} 
This research was supported in part by the Swiss National Science Foundation
(SNF) under contract 200020-109162, 
 by the UK Science and Technology Facilities Council and by the European 
Commission under contract MRTN-2006-035505 (Heptools).

%%%%%%%%%%%%%%%%%%%%%%%%%%%%%%%%%%%%%%%%%%%%%%%%%%%%%%%%%%%%%%%%%%%%%

\end{document}